\documentclass[twocolumn,superscriptaddress,showpacs,preprintnumbers,amsmath,amssymb,prb]{revtex4}
\usepackage{txfonts}

\usepackage{graphicx}
\usepackage{dcolumn}
\usepackage{bm}
\usepackage{amsmath,amssymb}

\begin{document}

\title{Analytical expression of geometrical pumping for a quantum dot \\based on quantum master equation}
\author{Ryosuke Yoshii} 
\affiliation{Yukawa Institute for Theoretical Physics, 
Kyoto University, Kitashirakawa Oiwake-Cho, Kyoto 606-8502, Japan}
\affiliation{Research and Education Center for Natural Sciences, 
Keio University, 4-1-1 Hiyoshi, Kanagawa 223-8521, Japan}

\author{Hisao Hayakawa}
\affiliation{Yukawa Institute for Theoretical Physics, Kyoto University, Kitashirakawa Oiwake-Cho, Kyoto 606-8502, Japan}

\date{\today}

\begin{abstract}
We analytically investigate a non-equilibrium quantum pumping 
for a single quantum dot connected to external leads 
on the basis of the quantum master equation (QME). 
We show that the Coulomb interaction associated with the spin effect in the dot induces the Berry-like phase in the parameter space and this phase results in the excess charge transfer for the cyclic modulation of parameters in leads. 
We obtain an analytical expression of the curvature of the phase and that for the pumped currents.   
\end{abstract}

\maketitle

\par
\section{Introduction}

The recent development in the nanotechnology enables us to explore the physics of 
quantum transport for highly tunable systems. 
In fact, one can make a nanometer scale confinement structure, known as the quantum dot (QD). 
Using attaching external leads can be attached via the tunnel coupling, electrons can transport through QD. 
There are a number of tunable parameters in the QDs, e.g., energy level in a QD, 
bias voltage, and tunnel barrier between the QD and the leads. 
Those tunablities enable us to compare the theory with the experiment for a variety of physical effects. 

Quantum transport phenomena attract theoretical and experimental investigations 
to reveal the quantum many-body properties under various non-equilibrium circumstances. 
One of the major issues of the quantum transport is the electron transfer induced by a cyclic temporal change of the parameters in 
the system. 
Those phenomena, known as quantum pumping, are widely observed. 
The electron pumping by the quantum effect has been proposed by Thouless.\cite{Thouless} 
This phenomenon is also closely connected with Berry phase and Berry curvature.\cite{Berry}  
In the cyclic modulation of parameter, 
the eigenstates can gain the Berry phase even if the Hamiltonian of the system itself return to the initial Hamiltonian. 
If the curvature corresponding to Berry phase is non zero in parameter space, 
the cyclic modulation of the parameter produces currents. 
The original idea of the quantum pumping for a closed system has been extended 
to an open system \cite{Buttiker, Buttiker2, Aleiner, Brouwer, Zhou, Andreev, Cremers, Moskalets, Stefanucci, Brouwer2, Breuer} 
and represented by a geometrical expression. \cite{Makhlin} 
Since then, the various effects on the quantum pumping have been investigated, e.g.\ Coulomb interaction in QD,\cite{Splettstoesser} interaction of two electrons in a triple-well structure.\cite{Tamborenea} 
The spin pumping have also been investigated in the presence of Kondo effects.\cite{Schiller, Aono}   
Experimentally, the quantum pumping has been realized by the transport experiment in the mesoscopic systems.\cite{Kouwenhoven, Pothier, Switkes, Buitelaar, Kaestner, Chorley} 
In those experiments, the quantized dc current has been obtained by modulation of the parameters.

The geometrical phase also appears in the context of master equation. 
If the parameter of the Liouvillian is adiabatically varied, 
the eigenstates obtain the geometrical phase similar to the Berry phase. 
This geometrical phase, so called Berry-Sinitsyn-Nemenman (BSN) phase, 
has originally been offered in the context of the classical master equation for stochastic systems.\cite{Parrondo, Usmani, Astumian, Sinitsyn1, Sinitsyn2, Rahav, Ren, SagawaHayakawa, Breuer} 
In these systems, the excess part of the cumulant generating function is 
expressed by a Berry-like phase on the eigenstate of the classical master equation. 
It is notable that Ref. \onlinecite{Ren} analyzed a spin-boson model under a cyclic modulation of two temperatures in the reservoirs, and obtained the pumped current. 
In Ref.\ \onlinecite{Yuge}, the QME for adiabatic modulation of the reservoir parameters has been analyzed, such as temperatures and chemical potentials in leads. 
The authors have demonstrated that the BSN phase exists in general, and thus, the pumped current can exist for general situations under the adiabatic modulations of reservoir parameters. 
They have applied their method to a double QDs system with inter-dot repulsion, 
and have obtained the BSN curvature for various interaction strength, which vanishes in no-interacting limit. 
It is remarkable that the quantum pumping is generated by the modulation of the parameters in thermal baths. 
This analysis has already been extended to demonstrate the existence of a path-dependent entropy in a nonequilibrium QD system.
\cite{Yuge2}
However, their calculation is numerical and, thus, explicit parameter dependence is unclear. 
Moreover, they have not taken into account the effect of spins in the QDs nor the precise intra-dot interaction. 
In this paper, thus, we analyze a more realistic system of the QD connected to external leads and obtain the BSN curvature and pumped current analytically. 
We show that the electron-electron interaction in QD induces the quantum pumping.

The present paper is organized as follows. 
In Sec.\ II, we introduce the model for the QD coupled to external leads and 
present a brief review of calculation for the pumped current in the basis of quantum master equation. 
In Sec.\ III, we apply the method presented in the Sec.\ II to our model. 
The analytical expression of BSN curvature for our model is derived. 
We show that the electron-electron interaction in QD induces the finite BSN curvature. 
We also show that we can obtain the pumped current by modulating the parameters in the bath.  
In Sec.\ IV, we summarize our results and discuss the implications for geometrical pumping under 
many body correlation. 
In Appendix A, we derive the QME for our model in the presence of counting field. 
In Appendix B, we present the detailed calculation for BSN curvature. 

\section{Model and method}
In this section, we present our model and the derivation for the geometrical expression of charge pumping and 
the corresponding curvature on the basis of QME. 
 
\subsection{Model}
\begin{figure}
\includegraphics[width=20pc]{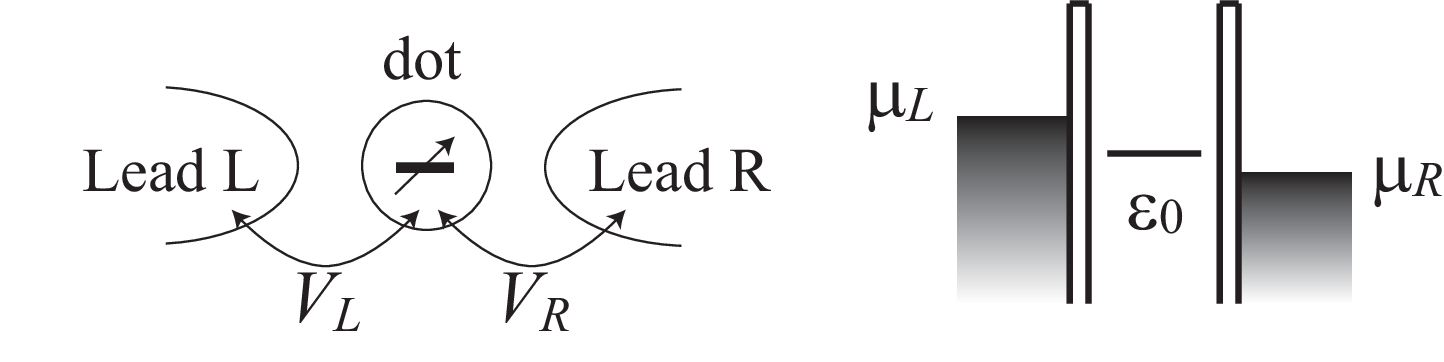}\hspace{2pc}%
\caption{A schematic picture of a quantum dot coupled with two leads. }
\label{model}
\end{figure}

We consider a QD connected with two external leads by using the Anderson model. 
Our model is depicted in Fig.\ {\ref{model}}. 
The Hamiltonian is given by
\begin{eqnarray}
H&=&H_{\mathrm{s}}+H_{\mathrm{bath}}+H_{\mathrm{int}},\label{H}\\
H_{\mathrm{s}}&=&\sum_{\sigma}\epsilon_0 d_\sigma^\dagger d_\sigma+Un_{\uparrow}n_{\downarrow},\label{H0}\\
H_{\mathrm{bath}}&=&\sum_{\gamma,k,\sigma}\epsilon_k a_{\gamma,k,\sigma}^\dagger 
a_{\gamma,k,\sigma},\label{H1}\\
H_{\mathrm{int}}&=&
\sum_{\gamma, k, \sigma}V_{\alpha} d^\dagger_{\sigma} a_{\alpha, k, \sigma}+\mathrm{h.c.},\label{H2}
\end{eqnarray}
where, $a_{\alpha, k,\sigma}^\dagger$ and $a_{\alpha, k,\sigma}$ are, respectively, the creation and the annihilation operators 
for the electron in the leads $\alpha$($=L,R$) with the wave number $k$, the energy $\epsilon_k$, and the spin $\sigma$.
Similarly, $d^\dagger_\sigma$ and $d_\sigma$ are those in the QD, respectively, and  
$n_{\sigma}=d^\dagger_{\sigma}d_{\sigma}$. 
$U$ and $V_\alpha$ are, respectively, the electron-electron interaction in the QD and the transfer energy between QD and the lead $\alpha$. 
Hereafter, we refer to the two leads as heat baths (HBs). 
We adopt a model in the wide band limit for leads, which corresponds to a boson system with Ohmic dissipation.
We denote, in this paper, the line width $\Gamma=\pi\nu V^2$ where $V^2=V_L^2+V_R^2$ and $\nu$ is the density of states in the leads. 

In this paper, we consider a geometrical pumping caused by an adiabatic modulation of the parameters in the HBs. 
In particular, 
we focus on the case that the chemical potential in left lead $\mu_L$ and right lead $\mu_R$ are adiabatically controlled 
throughout this paper. 
It is straightforward to expand the parameter space to include the other parameters, such as the temperatures in leads, 
the energy level of QD, the tunnel potential, and so on. 
In this paper we consider the case that the parameters except for the chemical potentials are symmetric in the left and the right leads, 
i.e.\ $T_L=T_R\equiv T$ and $V_L=V_R\equiv V$. 
The generalization toward asymmetric tunneling or asymmetric temperature case is 
straightforward and that adds no qualitative change to following arguments. 
For simplicity, we set $\hbar=1$, $e=1$ and $k_{\mathrm{B}}=1$.

We neglect the Kondo effect in this paper. 
This treatment can be justified at temperatures much higher than Kondo temperature $T_{\mathrm{K}}$. 
Because $T_{\mathrm{K}}$ is typically much smaller than the line width of the energy level in QD, 
the following argument is expected to be valid for wide range of temperatures.

\subsection{Method}
In Ref.\ \onlinecite{Yuge}, the excess charge transfer has been calculated on the basis of QME. 
This is a suitable way to express the charge pumping as the geometrical quantity. 
Since the charge pumping is represented as the Berry-like phase, 
the geometrical pumping in the system is characterized by the corresponding curvature (BSN curvature). 

First, we introduce the counting field $\chi$, in order to count the number of the electrons which transfer from the left lead to the right lead. 
Then the generation function of the number of the transferred electrons is given by 
\begin{equation}
Z(\chi)=\langle e^{-i\chi N_t}e^{i\chi N_0}\rangle,
\label{Gf}
\end{equation}
where $N_0$ and $N_t$ are the number of the electrons in left leads at initial time $0$ and 
at time $t$, respectively. 
The cumulant generating function (CGF) is given by $S(\chi)\equiv \ln Z(\chi)$ and 
 $\partial F(\chi)/\partial (i\chi)|_{\chi=0}$ yields $\langle N_t-N_0\rangle$. 
From now on, we assume that commutable relation $[H_s,N]=0$ holds. 
In other words, we decompose full Hamiltonian as $H_s+H_{\textrm{int}}$ as $[H_s,N]=0$ holds. 
Thus Eq.\ (\ref{Gf}) is rewritten as 
\begin{equation}
Z(\chi)=
\mathrm{Tr}
\left[
e^{iH_{\chi} t}
e^{-iH_{-\chi} t} 
\rho_0
\right],
\label{Gf2}
\end{equation}
where $\rho_0$ is the density matrix of the whole system in the initial state, and $H_{\pm \chi}$ satisfies
\begin{equation}
e^{iH_{\pm\chi} t}=e^{\pm\frac{i}{2}\chi N}
e^{iH t}
e^{\mp\frac{i}{2}\chi N}.
\end{equation}
Equation (\ref{Gf2}) describes the following ``time evolution".
The initial state described by $\rho_0$ evolves by the ``Hamiltonian" 
$H_{-\chi}$ from the initial time to $t$ and the state at $t$ evolves backward by $H_{\chi}$. 

In order to analyze Eq.\ (\ref{Gf2}), we make another assumption, corresponding to the Markovian approximation. 
In Eq.\ (\ref{Gf}), if $t$ is much larger than the tunnel rate $\sim \hbar /\Gamma$, the memory effect is negligible. 
Thus the time evolution of $\rho^\chi$ becomes local in time.  
By integrating out the degrees of freedom for conduction electrons, we obtain the following formal quantum master equation. 
\begin{equation}
\frac{d\rho^\chi}{dt}=\mathcal{K}^\chi\rho^{\chi}.
\label{QME}
\end{equation}
This is the generalization of the QME in the presence of counting field 
and thus we call this equation as the generalized QME (GQME). 
The explicit form of ${\cal K}^{\chi}$ corresponding to the setup in Fig.\ {\ref{model}} will be presented in the next section.

Next we consider the modulation of the parameters. 
For the consistency with the above discussion, the modulation must be slow compared with the tunnel rate. 
We follow the method described in Ref.\ \onlinecite{SagawaHayakawa}. 
The density matrix of the system can be expanded as 
\begin{equation}
\rho^\chi(t)=\sum_n c_n(t)e^{\Lambda^\chi_n(t)}\rho^\chi_n(\vec{\alpha}(t)),
\label{rhot}
\end{equation}
where $\vec {\alpha}(t)$ denotes the set of parameters at time $t$, 
$\lambda_n$ and $\rho^\chi_n$ are, respectively, the eigenvalue and the corresponding right eigenfunction 
of the $\mathcal{K}^\chi$ in Eq.\ (\ref{QME}). 
Here, $\Lambda_n^\chi(t)\equiv \int^t_0 dt^\prime\lambda^\chi_n(\vec \alpha(t^\prime))$ is the dynamical phase 
which corresponds to the house-keeping part of the generating function (\ref{Gf}). 
If $t$ is much larger than the characteristic time,
Eq.\ (\ref{rhot}) is reduced to
\begin{equation}
\rho^\chi(t)=c_0(t)e^{\Lambda^\chi_0(t)}\rho^\chi_0(\vec{\alpha}(t)),
\label{rhot2}
\end{equation}
where $\rho^\chi_0$ is the eigenvector whose eigenvalue $\chi_0$ has the maximum real part.
The order of the time gap between the longest and the second longest relaxation times for the eigenmodes can be estimated as $1/\textrm{Re} (\bar \lambda_0-\bar \lambda_1)$, 
where $\lambda_1$ has the second maximum real part and $\bar \lambda_n$ stands for the time average of $\lambda_n$. 
The above approximation corresponds to the adiabatic condition of the parameter modulation. 
Thus we obtain the excess part 
by subtracting the house-keeping part from the total CGF. 
The excess part of the CGF becomes  
\begin{equation}
S_{\mathrm{ex}}(i\chi)=-\int_{C}
l^\chi_{0,\vec{\alpha}(t)}
\cdot d\rho^\chi_{0,\vec{\alpha}(t)}+\mathrm{surface\ terms},
\label{Generatingfn}
\end{equation}
where $l^\chi_{0,\vec{\alpha}(t)}$, $d\rho_{0,\vec{\alpha}(t)}\equiv d\rho_{0,\vec{\alpha}(t)}^{\chi=0}$,  
is the left eigenvector corresponding to the eigenvalue $\lambda_0^\chi$. 
The total derivative on the right hand side of Eq.\ (\ref{Generatingfn}) is taken in the parameter space.
Thus we obtain the average number of electrons transfered from the left lead to the system 
by differentiating Eq.\ (\ref{Generatingfn}) with respect to $i\chi$, 
\begin{equation}
\Delta N=-\int_{C}
l^\prime_{0,\vec{\alpha}(t)}
\cdot 
d\rho_{0,\vec{\alpha}(t)},
\label{excessN}
\end{equation}
where $l^\prime_{0,\vec{\alpha}(t)}=\partial l^\chi_{0,\vec{\alpha}(t)}/\partial(i\chi)|_{\chi=0}$ 
and $C$ represent the integration path along the trajectory in parameter space, respectively.
Thus, $\Delta N$ is expressed as the integration of Berry-like phase. 
Indeed, the Berry-like phase has the corresponding BSN curvature, as 
\begin{equation}
F_{\mu_L,\mu_R}d\mu_L\wedge d\mu_R =
dl^\prime_{0,\vec{\alpha}(t)}\wedge
d\rho_{0,\vec{\alpha}(t)},
\label{excessNcurvature}
\end{equation}
where $\wedge$ is the wedge product. 
The above argument can be easily generalized to the higher cumelant which is related to the valiance, skewness, and so on.

\section{Results}
In this section, we apply the method described in the previous section to our model depicted in Fig.\ \ref{model}. 
In the first subsection, we derive the GQME for our system. 
In the second subsection, we calculate the BSN phase and curvature for our system. 
We also estimate the geometrical pumping for cyclic modulation.

\subsection{Quantum Master Equation}
In our system, the GQME becomes the rate equation between 
$\rho\equiv (\rho_d, \rho_{\uparrow}, \rho_{\downarrow}, \rho_e)^T$, corresponding to the 
double occupied state, singly occupied state (up spin or down spin), and empty state, respectively.
Indeed this is the generalization of the QME derived in Ref.\ \onlinecite{Thielmann}. 
The generalized Liuvillian for our system in a wide band case is 
\begin{eqnarray}
\mathcal{K}^\chi_{\vec\alpha}
=\mathcal{K}^{(L\rightarrow D)\ \chi}_{\vec\alpha}+\mathcal{K}^{(D\rightarrow L)\ \chi}_{\vec\alpha},\label{liouvillian}\\
\mathcal{K}^{(L\rightarrow D)\ \chi}_{\vec\alpha}=\Gamma\left(
\begin{array}{cccc}
0 &f_+^{(1)\chi}&f_+^{(1)\chi}&0\\
0&-f_+^{(1)} &0&f_+^{(0)\chi}\\
0&0&-f_+^{(1)}&f_+^{(0)\chi}\\
0&0&0&-2f_+^{(0)}
\end{array}
\right),\label{KLD}\\ 
\mathcal{K}^{(D\rightarrow L)\ \chi}_{\vec\alpha}=\Gamma\left(
\begin{array}{cccc}
-2f_-^{(1)} &0&0&0\\
f_-^{(1)\chi}&-f_-^{(0)}&0&0\\
f_-^{(1)\chi}&0&-f_-^{(0)}&0\\
0&f_-^{(0)\chi}&f_-^{(0)\chi}&0
\end{array}
\right),\label{KDL}
\end{eqnarray}
in the leading order of $V$ with Markovian approximation. (For derivation, see Appendix A.) 
Here, $f_+^{(j)\chi}= e^{i\chi} f_{L}^{(j)}(\epsilon_0)+f_R^{(j)}(\epsilon_0)$ 
and $f_-^{(j)\chi}= e^{-i\chi} [1-f_L^{(j)}(\epsilon_0)]+1-f_R^{(j)}(\epsilon_0)$ 
where we have introduced Fermi distribution function $f_\gamma^{(j)}(\epsilon_0)=\left[1+e^{-\beta(\epsilon_0+jU-\mu_\gamma)}\right]^{-1}$ 
in the lead $\gamma$ with the inverse temperature 
$\beta=T^{-1}$. 
Note that $f_{\pm}^{(j)}$ satisfies $f_{\pm}^{(j)}\equiv f_{\pm}^{(j) \chi=0}$. 
The Liouvillian $R^{(L\rightarrow D)\ \chi}_{\vec\alpha}$ represents 
the process that the electron transfers from the leads to the QD 
and $R^{(D\rightarrow L)\ \chi}_{\vec\alpha}$ represent the opposite process. 
As we mentioned, the parameters are denoted by a vector $\vec{\alpha}(t)=(\mu_L, \mu_R)$, 
which is easy to control in an experiment. 
The following calculation can be extended to the other parameter spaces.

\subsection{BSN curvature and pumped current}
\begin{figure}
\includegraphics[width=20pc]{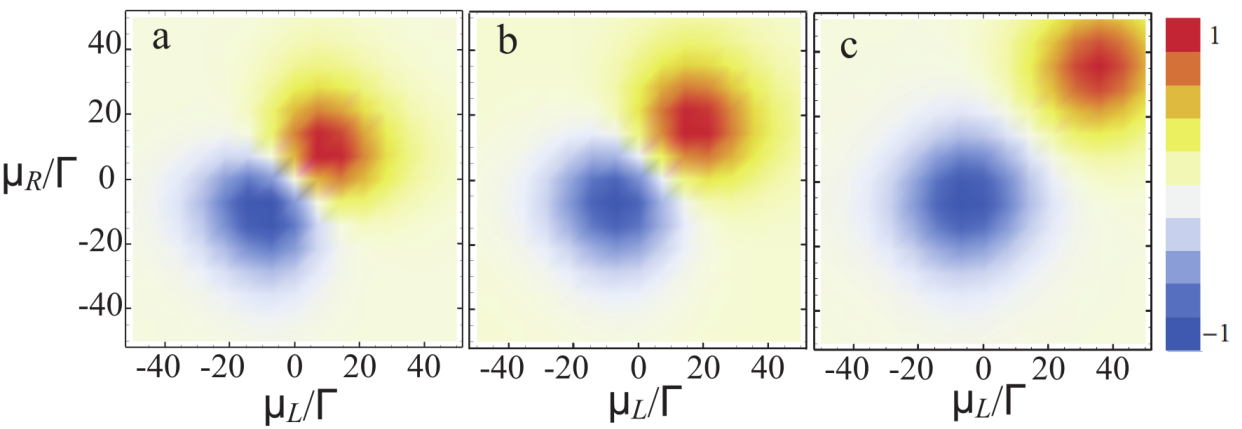}
\caption{(Color online) The BSN curvature in $\mu_L$-$\mu_R$ plane for $U=\Gamma$(Fig.\ a), $U=20\Gamma$(Fig.\ b), and $U=40\Gamma$(Fig.\ c), where temperature is set to be $T=10\Gamma$.  
We normalize the peak and dip values to be $+1$(Red) and $-1$(Blue), 
though those values depend on $U$. (See Figs.\ 3 and 4.) 
The peak and dip positions lie near $\epsilon_0$ and $\epsilon_0+U$, respectively.
The half width is approximately equal to $\sim T$ for each case.}
\label{BSNcurvature}
\end{figure}

\begin{figure}
\includegraphics[width=15pc]{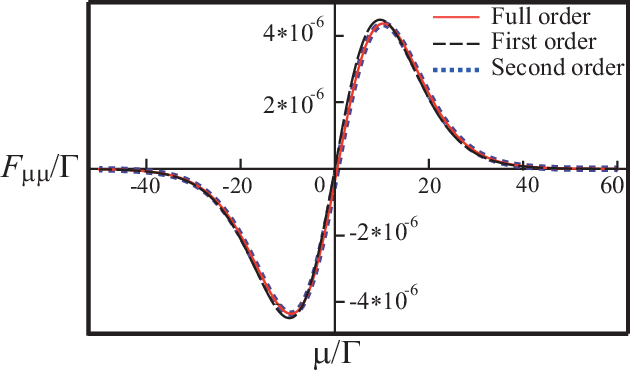}\hspace{2pc}%
\caption{The peak structure of BSN curvature along $\mu_L=\mu_R$ line for $T=10\Gamma $ and $U=\Gamma$.  
This result corresponds to the case of $U\ll T$. 
Here, the solid line shows the full result of the adiabatic expression of  Eqs. (\ref{QME}), (\ref{rhot2}) and (\ref{liouvillian})-(\ref{KDL}) obtained by the numerical calculation, 
while the broken line and dotted line illustrate the results of the first order and second order perturbation, respectively. }
\label{peak-1}
\end{figure}

In this subsection, we analyze the BSN curvature for two limiting cases, 
$U/\Gamma \ll T/\Gamma$ or $U/\Gamma \gg T/\Gamma$. 
First, we show that the Coulomb interaction induces the BSN curvature. 
This means that, the BSN curvature is zero for $U=0$. 
In the case of $U=0$, the density matrix of the system can be decomposed into $\uparrow$ spin space and $\downarrow$ spin space 
and thus the model is equivalent to the two noninteracting spinless Fermion system. 
The Generalized Liuvillian for each spin becomes 
\begin{eqnarray}
\mathcal{K}^{\chi}_{\vec\alpha}=\Gamma\left(
\begin{array}{cccc}
-f_-^{(0)} &f_+^{(0)\chi}\\
f_-^{(0)\chi}&-f_+^{(0)}
\end{array}
\right).
\label{GLiou0}
\end{eqnarray}
Then we can easily verify that the BSN curvature is zero from Eq.\ (\ref{excessNcurvature}) (see Appendix B).
The BSN curvature for $U\ll T$ is given by the expansion of  $\beta U$ as
\begin{eqnarray}
F_{\mu_L, \mu_R}=\frac{\beta^2}{(2f_+ +f_-)^3}\left(A_1\beta U+\cdots\right),
\end{eqnarray}
where
\begin{eqnarray}
A_1=-\frac{e^{2\beta\epsilon_0}-e^{\beta(\mu_L+\mu_R)}}
{(e^{\beta\epsilon_0}+e^{\beta\mu_L})^3(e^{\beta\epsilon_0}+e^{\beta\mu_R})^3}
e^{\beta(2\epsilon_0+\mu_L+\mu_R)},\ \cdots.
\label{A1}
\end{eqnarray}
The result clearly exhibits that the geometrical pumping is forbidden in following situations, 
(i) at high temperature limit and 
(ii)`` macroscopic" modulation $|\epsilon_0-\mu_L|, |\epsilon_0-\mu_R| \gg U$. 
In the situation (ii), the separation between the peak and the dip for the BSN curvature becomes smaller than the energy scales 
$|\epsilon_0-\mu_L|$ and $|\epsilon_0-\mu_R|$. 
Then the effect from the peak and that from the dip are canceled out. 
Thus we conclude the geometrical pumping is purely quantum effect, which is contrast to the case for the spin-boson model, equivalent to a classical two-level system. 
In Fig.\ \ref{BSNcurvature}, we plot the BSN curvature for various strength $U$. 
This result illustrates that $U$ makes the peak and the dip separate. 
In Fig.\ \ref{peak-1}, the peak position and the magnitude along $\mu_L=\mu_R$ line are plotted. 
The first order expansion yields good approximation and second order expansion is indistinguishable from the full numerical calculation of the adiabatic model.
This figure , thus, ensures that the perturbation treatment is valid.

\begin{figure}
\includegraphics[width=15pc]{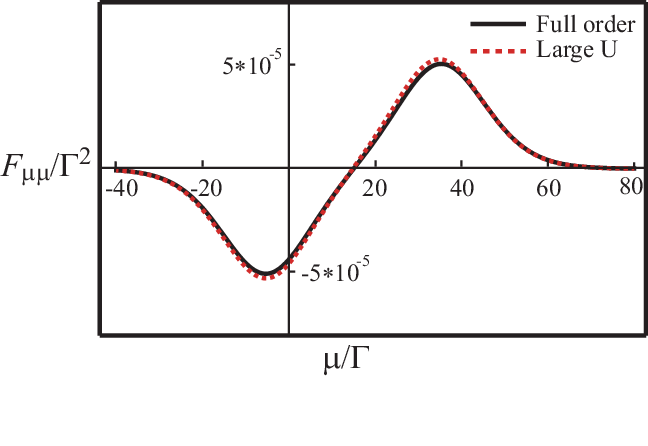}\hspace{2pc}%
\caption{The peak structure of BSN curvature along $\mu_L=\mu_R$ line for $T=10\Gamma$. 
The solid line shows the full expression obtained by the numerical calculation for $U=40\Gamma$. 
The dotted line shows the result obtained by adopting the analytical expression for $U/T\gg 1$ to $U=40\Gamma$.
In this case, though $U$ is not significantly large ($U/T=4$), 
we obtain a good coincidence between the analytical expression for large $U$ and the full calculation of the adiabatic expression.  }
\label{peak-2}
\end{figure}

In real experimental setups we can assume the situation of $U/\Gamma\gg 1$ and $U/T\gg 1$ where the double occupied state is forbidden 
when $\mathrm{max}\{|\epsilon_0-\mu_L|, |\epsilon_0-\mu_R|\} \ll \mathrm{min}\{|U-\mu_L|, |U-\mu_R|\}$.
Thus, the basis of the GQME reduces to $\tilde \rho=(\rho_\uparrow, \rho_\downarrow, \rho_e)$ and generalized Liuvillian becomes\cite{Utsumi} 
\begin{eqnarray}
\mathcal{K}^\chi_{\vec \alpha}=\left(
\begin{array}{ccc}
-f_-^{(0)}(\epsilon_0)&0&f_+^{(0)\chi}(\epsilon_0)\\
0&-f_-^{(0)}(\epsilon_0)&f_+^{(0)\chi}(\epsilon_0)\\
f_-^{(0)\chi}(\epsilon_0)&f_-^{(0)\chi}(\epsilon_0)&-2f_+^{(0)}(\epsilon_0)
\end{array}
\right).
\label{GLiou}
\end{eqnarray}
In the case, the pumped current and the BSN curvature are, respectively, given by
\begin{equation}
\Delta N=-2\int_C \frac{1+f_L^{(0)}(\epsilon_0)}{2f_+^{(0)}(\epsilon_0)+f_-^{(0)}(\epsilon_0)}
dg,
\label{pumpN}
\end{equation} 
where $g=f_+^{(0)}(\epsilon_0)/[2f_+^{(0)}(\epsilon_0)+f_-^{(0)}(\epsilon_0)]$ and
\begin{equation}
F_{\mu_L\mu_R}=\frac{4}{(2f_+^{(0)}+f_-^{(0)})^3}
\frac{\partial f_L^{(0)}(\epsilon_0)}{\partial\mu_L}\frac{\partial f_R^{(0)}(\epsilon_0)}{\partial\mu_R}.
\label{Uinf}
\end{equation}
This expression clearly shows that the peak of the BSN curvature is near $(\mu_L, \mu_R)=(0, 0)$. 
We can also calculate the BSN curvature for the case of 
$U/T\gg 1$ and $\mathrm{min}\left\{|\epsilon_0-\mu_L|, |\epsilon_0-\mu_R|\right\} \gg \mathrm{max}\left\{|U-\mu_L|, |U-\mu_R|\right\}$.
In this case, the BSN curvature becomes
\begin{equation}
F_{\mu_L\mu_R}=-\frac{4}{(2f_-^{(1)}+f_+^{(1)})^3}
\frac{\partial f_L^{(1)}(\epsilon_0)}{\partial\mu_L}\frac{\partial f_R^{(1)}(\epsilon_0)}{\partial\mu_R}.
\label{Uinf2}
\end{equation}
Thus in the case of $U/T \gg 1$, the BSN curvature has a peak around  $(\mu_L, \mu_R)=(0, 0)$ and a dip around $(\mu_L, \mu_R)=(U, U)$.
We plot the BSN curvature along $\mu_L=\mu_R$ line in Fig.\ \ref{peak-2}. 
It is shown that the full adiabatic expression of the BSN curvature 
is well approximated by a superposition of Eqs.\ (\ref{Uinf}) and (\ref{Uinf2}) even when $U=40\Gamma$ ($U/T=4$).
For $T\ll \Gamma$, Eq.\ (\ref{Uinf}) becomes 
\begin{equation}
F_{\mu_L\mu_R}\simeq \frac{4}{(2f_+^{(0)}+f_-^{(0)})^3}
\delta (\mu_L-\epsilon_0)\delta (\mu_R-\epsilon_0).
\end{equation}
Thanks to the Stokes theorem
\begin{equation}
\Delta N=\int_S dS F_{\mu_L\mu_R},
\label{pumpN2}
\end{equation} 
the pumped current is given by 
\begin{equation}
\Delta N=\left.\frac{4}{(2f_+^{(0)}+f_-^{(0)})^3}\right|_{\epsilon_0=\mu_L,\epsilon_0=\mu_R},
\end{equation}
where $\int_S$ denotes the integration over the region enclosed by the trajectory of the parameter modulation. 
Then we can conclude we achieve the charge pumping of order $\sim e$ in one cycle. 

\begin{figure}
\includegraphics[width=15pc]{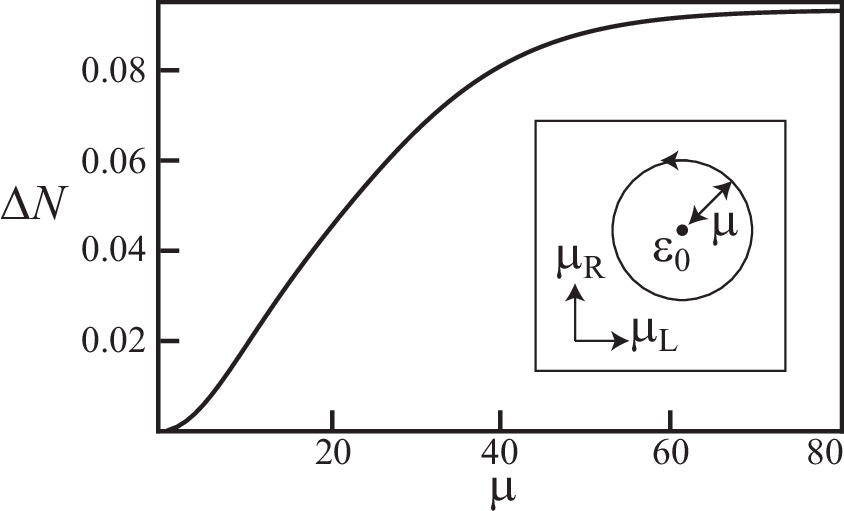}\hspace{2pc}%
\caption{The current obtained by cyclic modulation in the parameter space. 
The trajectory of the parameter modulation is depicted in inset. 
We make the cyclic modulation of radius $\mu$ from the energy level in QD. 
We use the analytical expression for $U\gg T$ and set $T=10\Gamma$. }
\label{currentcircle}
\end{figure}

We also calculate the pumped current at high temperatures $T=10\Gamma$ (see Fig.\ \ref{currentcircle}). 
We plot the pumped current against the radius of parameter modulation. 
We set the origin of the circle as $\epsilon_0$($=0$).
Even in the case of high temperatures, we can obtain the charge pumping of order $\sim e$ in $\sim 10$ cycles of the parameter modulation. 

\section{Conclusion and discussion}
We have analyzed the geometrical pumping for a system of a QD connected with two external leads based on the QME approach. 
We have derived the analytical expressions for BSN curvature for the case of $\beta U\ll 1$ or $\beta U\gg 1$. 
We have shown that the Coulomb interaction in the QD causes the non-zero BSN curvature in $\mu_L$-$\mu_R$ plane. 
For $U=0$, the BSN curvature is zero in $\mu_L$-$\mu_R$ plane 
and the peak and the dip emerge for $U\ne 0$. 
In contrast to the spin boson model and the spinless model,\cite{Esposito, Yuge} 
the electron-electron interaction in the QD has the central role in this system. 
We also analyze the BSN curvature when $\epsilon_0$ and $\epsilon_0+U$ are well separated, 
where the peak and the dip of the curvature appear near $\epsilon_0$ and $\epsilon_0+U$, respectively. 
Thus the quantum pumping can be achieved in our setup with parameter modulation only in the HBs. 

Throughout this paper, we have ignored the Kondo correlation which is equivalent to the case $T\gg T_{\mathrm{K}}$. 
The examination of strong coupling regimes ($T\ll T_{\mathrm{K}}$) would require calculations 
using the renormalization improved perturbation theory, self consistent Born approximation, 
etc., which is beyond the scope of the present paper. 
The investigation of the finite time modulation which brings non-adiabatic correction and non-Markovian effect 
are also the future issues.  
In the case of boson transport, the non-adiabatic effect have been studied and 
the generalization to the case of fermion transport is expected to be straightforward.\cite{Watanabe}

\section*{Acknowledgment}
The authors acknowledge T.\ Sagawa and T.\ Yuge for their helpful advices and K.\ Watanabe for valuable discussion. 
R.\ Y.\ is the Yukawa Fellow and this work is partially supported by Yukawa Memorial Foundation.

\appendix
\section{GQME for Anderson model}
In this appendix we summarize the derivation of Eqs.\ (\ref{liouvillian}) - (\ref{KDL}).
As is assumed, the density matrix of full system $\rho$ is decomposed into the matrix of system $\rho_{\mathrm{S}} $and 
the matrix of the bath in thermal equilibrium $\rho_{\textrm{bath}}$ at initial time $t_0$,
\begin{equation}
\rho(t_0)=\rho_{\mathrm{s}}(t_0)\rho_{\mathrm{bath}}.
\label{initrho}
\end{equation}
The time evolution of $\rho(t)$ is described by Liouvillian as 
\begin{equation}
\frac{d}{dt}\rho(t)=\mathcal{K}\rho(t).
\label{QME0}
\end{equation}
As is the case of Hamiltonian (\ref{H}), the full Liouvillian $\mathcal{K}$ can be decomposed into $\mathcal{K}_{\mathrm{s}}$, $\mathcal{K}_{\mathrm{bath}}$, and $\mathcal{K}_{\mathrm{int}}$. 
As can be seen from the form of time evolution equation, the Dyson's equation in quantum system can be used. 
By the Laplace transformation of $\rho(t)$ 
\begin{equation}
\rho(z)=\int^\infty_{t_0}dt e^{-z(t-t_0)}\rho(t),
\end{equation}
the Eq.\ (\ref{QME}) becomes 
\begin{equation}
\rho(z)=\frac{1}{z-\mathcal{K}}\rho(t_0)=G^0(z)+G^0(z)\mathcal{K}_{\mathrm{int}}G^0(z)+\cdots,
\label{QMELspace}
\end{equation}
where $G^0(z)=(z-\mathcal{K}_{\mathrm{s}}-\mathcal{K}_{\mathrm{bath}})^{-1}$. 
From Eqs.\ (\ref{initrho}) and (\ref{QMELspace}), 
the reduced density matrix which is obtained by tracing out the bath degrees of freedom becomes 
\begin{equation}
\rho_\mathrm{s}(z)=\mathrm{Tr}_{\mathrm{bath}}\left[\left(G^0(z)+
G^0(z)\mathcal{K}_{\mathrm{int}}G^0(z)\mathcal{K}_{\mathrm{int}}G^0(z)\right)\rho_s (t_0)\rho_{\mathrm{bath}}\right],
\label{rhosz}
\end{equation}
in the second order of $\mathcal{K}_{\mathrm{int}}$. 
It can be shown that the term linear to $\mathcal{K}_{\mathrm{int}}$ vanishes. 
By using $\mathcal{K}_{\mathrm{bath}}\rho_{\mathrm{bath}}=0$, the first term of Eq.\ (\ref{rhosz}) can be rewritten as 
\begin{equation}
\mathrm{Tr}_{\mathrm{bath}}\left[G^0(z)\rho_s (t_0)\rho_{\mathrm{bath}}\right]=
G^0_{\mathrm{s}}(z)\rho_{\mathrm{s}}(t_0),
\label{Tr1st}
\end{equation}
where $G^0_{\mathrm{s}}(z)=(z-\mathcal{K}_{\mathrm{s}})^{-1}$. 
Now we define the effective Liouvillian $\mathcal{K}_\mathrm{eff}$, 
which describes the time evolution of $\rho_\mathrm{s}$ as 
\begin{equation}
\rho_{\mathrm{s}}(z)=\frac{1}{z-\mathcal{K}_\mathrm{eff}(z)}\rho_{\mathrm{s}}(t_0).
\label{rho_sTE}
\end{equation}
This is equivalent to the following time evolution equation 
\begin{equation}
\frac{d}{dt}\rho_{\mathrm{s}}=\int_{t_0}^td\tau \mathcal{K}_{\mathrm{eff}}(t-\tau)\rho_{\mathrm{s}}(\tau).
\label{terhoeff}
\end{equation}
Here we can see the non-Markovian effect, i.e.\ memory effect from previous times. 
By decomposing $\mathcal{K}_{\mathrm{eff}}(z)$ into the ``free part" and the ``self energy part" as 
$\mathcal{K}_{\mathrm{eff}}=\mathcal{K}_{\mathrm{s}}+\Sigma(z) $, 
it becomes more clear that the memory effect is induced by the interaction. 
By expanding (\ref{rho_sTE}) in $\mathcal{K}_\mathrm{eff}(z)$, we obtain 
\begin{equation}
\rho_{\mathrm{s}}(z)=\left(G^0_{\mathrm{s}}(z)+
G^0_{\mathrm{s}}(z)\Sigma (z) G^0_{\mathrm{s}}(z)+\cdots \right)\rho_s (t_0),
\label{rhoeffz}
\end{equation}
where $G^0_{\mathrm{s}}(z)=(z-\mathcal{K}_{\mathrm{s}})^{-1}$. 
From Eqs.\ (\ref{rhosz}), (\ref{Tr1st}), and (\ref{rhoeffz}), we can easily see that the second order term in Eq.\ (\ref{rhosz}) 
is equal to second term in Eq.\ (\ref{rhoeffz}). 
Thus the lengthy calculation yields\cite{Leijnse} 
\begin{equation}
\Sigma(z)=- \sum_{c,c^\prime=\pm } \sum_{\xi=\pm}c c^{\prime} J^{c^\prime}_{-\xi, \sigma}  
|aa^\prime\rangle \rangle \langle\langle aa^\prime|
 J^{c}_{\xi, \sigma^\prime} I(\xi, c, a, a^\prime), 
\end{equation}
where $|a, b \rangle\rangle =|a\rangle \langle b| $ is the two state vector and $J^{c}_{\xi, \uparrow}$ is the ladder operators (for the case of $\downarrow$, the definition is the same) defined as 
\begin{eqnarray}
&&J^{+}_{+, \uparrow}=|e, \alpha \rangle\rangle\langle\langle \uparrow, a | 
+ |\downarrow, \alpha \rangle\rangle\langle\langle d, a |, \\
&&J^{+}_{-, \uparrow}=|\uparrow, \alpha \rangle\rangle\langle\langle e, a | 
+ |\downarrow, \alpha \rangle\rangle\langle\langle d, a |, \\
&&J^{-}_{+, \uparrow}=|\alpha, \uparrow \rangle\rangle\langle\langle a, e | 
+ |\alpha, d \rangle\rangle\langle\langle a, \downarrow |, \\
&&J^{-}_{-, \uparrow}=|\alpha, e \rangle\rangle\langle\langle a, \uparrow | 
+ |\alpha, \downarrow \rangle\rangle\langle\langle a, d |, 
\end{eqnarray}
and $I(\xi, c, a, a^\prime)$ is given by 
\begin{equation}
I=V^2\sum_k \left(\frac{f_\gamma^{-\xi c} (\omega_k)}{z+i\xi \omega_{k}+i\Delta_{a,a^\prime}}\right), 
\label{spectrum}
\end{equation}
where $\Delta_{a,a^\prime}=\epsilon_a-\epsilon_{a^\prime}$ is the energy difference of the QD states. 
In the case of flat band $d \omega_k /d k=\mathrm{const}$, the function $I$ which describes the effect of spectrum for the leads on tunneling process, can be rewritten as 
\begin{equation}
I=\frac{1}{\pi} \Gamma \int ^{D}_{-D} d\omega \frac{f_\gamma^{-\xi c} (\omega)}{z+i\xi \omega+i\Delta_{a,a^\prime}},
\end{equation}
where the line width $\Gamma$ is defined as $\Gamma =\pi \nu V^2$ with the density of states in leads $\nu$. 

Next we make an assumption which corresponds to neglect the memory effect in Eq.\ (\ref{terhoeff}). 
This is valid when the time scale of the dynamics of the system is much larger than that of bath. 
By taking the long-time limit $z\rightarrow +0$, 
we can use the relation $\lim_{\eta\rightarrow +0} (\omega+i\eta)^{-1}=-i\pi \delta(\omega)+P \omega^{-1}$. 
Assuming the wide band limit $D\rightarrow \infty$, the imaginary part of $\Sigma$ can be negligible and thus, we obtain 
\begin{equation}
\Sigma(z)=-\frac{1}{2}\Gamma  \sum_{c,c^\prime=\pm } \sum_{\xi=\pm}c c^{\prime} J^{c^\prime}_{-\xi, \sigma}  |a a^\prime \rangle \rangle \langle\langle aa^\prime|
 J^{c}_{\xi, \sigma^\prime} f_\gamma^{-c\xi }(\Delta_{aa^\prime}). 
\end{equation}
By replacing $\Sigma(z)\rightarrow \Sigma(+0)$ (long-time limit),  Eq.\ (\ref{terhoeff}) can be rewritten as
\begin{equation}
\frac{d}{dt}\rho_{\mathrm{s}}=\mathcal{K}_{\mathrm{eff}}(z=+0) \rho_{\mathrm{s}}(t).
\end{equation}
Thus the Liouvillian for our system without the counting field becomes Eq.\ (\ref{liouvillian}) with $\chi=0$. 
In the present case, the counting field can be easily taken into account by the transposition, 
$a_{L, \sigma}^{\dagger}\rightarrow e^{-i\chi/2}a_{L, \sigma}^{\dagger}$ and $a_{L, \sigma}\rightarrow e^{i\chi/2}a_{L, \sigma}$.\cite{Bagrets} By using this method, we arrive at Eq.\ (\ref{liouvillian}).

\section{Detailed calculation for BSN curvature}
\subsection{In the case of $\beta U\ll 1$}
When $U=0$, the density matrix of the system can be decomposed into $\uparrow$ spin space and $\downarrow$ spin space 
and the Generalized Liuvillian for each spin is given by Eq.\ (\ref{GLiou0}). 
The eigenvalues of the matrix are 
\begin{equation}
\lambda_{\pm}=\Gamma\left[-1\pm\sqrt{D^{\chi}}\right],
\end{equation}
with $D^\chi=1+(f_+^{(0)\chi}f_-^{(0)\chi}-f_+^{(0)}f_-^{(0)})$, where $\lambda_+$ has the maximum real part.
Thus the right eigenstate corresponding to $\lambda_+$ for $\chi=0$ is  
\begin{equation}
\rho^{\chi=0}_0=\frac{1}{2}\left(f_+^{(0)}, f_-^{(0)}\right)^{T}
\label{rho22}
\end{equation}
and left eigenstate corresponding to $\lambda_+$ and its derivative with $\chi$ at $\chi=0$ is 
\begin{eqnarray}
&&l^{\chi}_0=\left(1, -\frac{1}{2f_-^{(0)\chi}}\left(f_+^{(0)}-f_-^{(0)}-\sqrt{D^\chi}\right)\right),\\
&&l^\prime_0=(0, 1),
\label{lprime22}
\end{eqnarray}
where $l^\chi_0$ is normalized to be $l^{\chi=0}_0=(1,1)$. 
Substituting Eqs.\ (\ref{rho22}) and (\ref{lprime22}) into Eq.\ (\ref{excessNcurvature}), 
we obtain $F_{\mu_L, \mu_R}=0$ for $U=0$. 

Next, we adopt the perturbation in terms of $\beta U$. 
The Liouvillian can be expanded in powers of $\beta U$, as 
\begin{eqnarray}
&&\mathcal{K}^\chi_{\vec\alpha}
=\left.\mathcal{K}^\chi_{\vec\alpha}\right|_{\beta U=0}+\beta U\left.\frac{\partial\mathcal{K}^\chi_{\vec\alpha}}{\partial \beta U} \right|_{\beta U=0}+\cdots, \label{uexpansion}\\
&&\left.\mathcal{K}^{\chi}_{\vec\alpha}\right|_{\beta U=0}=\Gamma\left(
\begin{array}{cccc}
-2f_- &f_+^{\chi}&f_+^{\chi}&0\\
f_-^{\chi}&-f_--f_+ &0&f_+^{\chi}\\
f_-^{\chi}&0&-f_--f_+&f_+^{\chi}\\
0&f_-^{\chi}&f_-^{\chi}&-2f_+
\end{array}
\right),\label{unperturbed}\\
&&\left.\frac{\partial \mathcal{K}^{\chi}_{\vec\alpha}}{\partial \beta U}\right|_{\beta U=0}=\Gamma\left(
\begin{array}{cccc}
-2\dot f_- &\dot f_+&\dot f_+&0\\
\dot f_-&-\dot f_+ &0&0\\
\dot f_-&0&-\dot f_+ &0\\
0&0&0&0
\end{array}
\right),\label{perturbed}
\end{eqnarray}
where we drop the superscript of $f_{\pm}$, and denote $(\partial f^{(1)} /\partial \beta U)_{\beta U=0}$ as $\dot f$. 
We treat the first term on the right hand side of Eq.\ (\ref{uexpansion}) as an unperturbed part. 
The four eigenstates of unperturbed part can be easily obtained from the tenser product of eigenstates for Eq.\ (\ref{GLiou0}). 
For example, the right eigenvectors can be written as  
\begin{equation}
\rho_{i,j}=(\rho^{(1)\uparrow}_{i} \rho^{(1)\downarrow}_{j},\ \rho^{(1)\uparrow}_{i} \rho^{(2)\downarrow}_{j},\ \rho^{(2)\uparrow}_{i} \rho^{(1)\downarrow}_{j},\ \rho^{(2)\uparrow}_{i} \rho^{(2)\downarrow}_{j})^T,\ (i, j=0, 1), 
\end{equation}
where $\rho_{0}$ is given by Eq.\ (\ref {rho22}) and $\rho_{1}$ is the right eigenstate corresponding to $\lambda_-$. 
Here we put superscript $\uparrow$ or $\downarrow$ to indicate the spin degrees of freedom and (i) denotes i-th component of density matrix vector. 
We can show that $\rho_{i,j}$ is indeed the eigenvectors of Eq.\ (\ref{unperturbed}), by decomposing to spin $\uparrow$ space and spin $\downarrow$ space 
\begin{eqnarray}
\left.\mathcal{K}^{\chi}_{\vec\alpha}\right|_{\beta U=0}=&&\Gamma\left(
\begin{array}{cccc}
-f_- &0&f_+^{\chi}&0\\
0&-f_- &0&f_+^{\chi}\\
f_-^{\chi}&0&-f_+&0\\
0&f_-^{\chi}&0&-f_+
\end{array}
\right)\nonumber\\
&&+\Gamma\left(
\begin{array}{cccc}
-f_- &f_+^{\chi}&0&0\\
f_-^{\chi}&-f_+ &0&0\\
0&0&-f_-&f_+^{\chi}\\
0&0&f_-^{\chi}&-f_+
\end{array}
\right).
\end{eqnarray}
The eigenstates $\rho_{00}$, $\rho_{01}$, $\rho_{10}$, and $\rho_{11}$ have corresponding eigenvalues $2\lambda_+$, $\lambda_++\lambda_-$, $\lambda_-+\lambda_+$, and $2\lambda_-$, respectively. 
Thus the eigenvector $\rho_{0,0}$ corresponds to the eigenvalue which has maximum real part. 
The left eigenstates can be obtained by the same procedure. 

For simplicity, we define $\beta U=\delta$ and denote Liouvillian as $K=K_0+\delta K_1+O(\delta^2)$.
We use the perturbation method.
First we expand the eigenvector and eigenvalue in $\beta$ as 
\begin{eqnarray}
&&R =\rho_{00}+\delta R_1+O(\delta^2), \\
&&\lambda=\lambda_++\delta \lambda_1+O(\delta^2), 
\end{eqnarray}
where $R$ is the eigenvector of $K$ with eigenvalue.$\lambda$. 
Here we show the calculation up to first order in $\delta$.
The left hand side and right hand side of $K R=\lambda R$ becomes 
\begin{eqnarray}
&&K R=\left(K_0+\delta K_1\right)\left(\rho_{00}+\delta R_1\right),\label{lhs}\\
&&\lambda R=\left(\lambda_++\delta \lambda_1\right)\left(\rho_{00}+\delta R_1\right)\label{rhs}.
\end{eqnarray}
By taking inner product with $\langle l_{00}|$, Eqs.\ (\ref{lhs}) and (\ref{rhs}) yield 
\begin{equation}
\lambda_1=\frac{ l_{00} K_1 \rho_{00}}{l_{00}\cdot\rho_{00}}.
\end{equation} 
Next we expand $R_1$ by eigenvectors $\rho_{i,j}$ as  
\begin{equation}
R_1=\sum_{(i,j)\neq (0,0)}C_{i,j} \rho_{i,j}.
\end{equation}
The Eqs.\ (\ref{lhs}) and (\ref{rhs}) lead to
\begin{equation}
C_{i,j}=\frac{ l_{i,j} K_1\rho_{00}}{2\lambda_+-\lambda_{i,j}},
\end{equation}
where $\lambda_{i,j}$ is the eigenvalue corresponding to $ l_{i,j}$. 
The resulting expression of $R_1$ becomes
\begin{eqnarray}
R_1=\frac{f_+\dot f_+}{8}
\left(
\begin{array}{c}
2+f_+\\
-f_+ \\
-f_+ \\
-2+f_+
\end{array}
\right).
\end{eqnarray}
From the same procedure, it is shown that the left eigenvector for $\chi=0$ is $(1,1,1,1)$ up to $O\left((\beta U)^2\right)$, 
namely the coefficient of the first order in $\beta U$ vanishes. 
We further apply the perturbation for $i\chi$ to $l^\chi$ and obtain 
\begin{eqnarray}
l^\prime &=&
-\frac{1}{2}\left(
\begin{array}{cccc}
2(f_+\dot f_+)^\prime, &
\left[\dot f_+ (f_+-1)\right]^\prime, &
\left[\dot f_+ (f_+-1)\right]^\prime, & 
0
\end{array}
\right)\nonumber \\
&&+O((\beta U)^2),
\end{eqnarray}
where $f^\prime =\left.\partial f/\partial (i\chi)\right|_{\chi=0}$. 
Thus the BSN curvature (\ref{excessNcurvature}) becomes Eq.\ (\ref{A1}).

\subsection{In the case of $\beta U\gg 1$}
In the case of $\beta U \gg 1$, the generalized Liouvillian for 
$|\epsilon - \mu_L|, |\epsilon - \mu_R| \ll |\epsilon+U - \mu_L|, |\epsilon+U - \mu_R|$ is 
given by Eq.\ (\ref{GLiou}). The eigenvalues of the matrix are 
\begin{equation}
\lambda_{0, \pm}=\Gamma f_-^{(0)},\ \frac{\Gamma}{2}\left[-(2f_+^{(0)}+f_-^{(0)})\pm\sqrt{\tilde D^{\chi}}\right],
\end{equation}
with $\tilde D^\chi=(2f_+^{(0)}+f_-^{(0)})^2+8(f_+^{(0)\chi}f_-^{(0)\chi}-f_+^{(0)}f_-^{(0)})$, where $\lambda_+$ has the maximum real part.
Thus the right eigenstate corresponding to $\lambda_+$ for $\chi=0$ is  
\begin{equation}
\rho^{\chi=0}_0=\frac{1}{2f_+^{(0)}+f_-^{(0)}}\left( f_+^{(0)}, f_+^{(0)}, f_-^{(0)}\right)^{T}
\label{rho33}
\end{equation}
and left eigenstate corresponding to $\lambda_+$ is
\begin{eqnarray}
&&l^{\chi}_0=\frac{1}{C^\chi}\left(-2f_-^{(0)}, -2f_-^{(0)}, C^\chi\right),
\label{lprime33}
\end{eqnarray}
with $C^\chi=2f_+^{(0)}-f_-^{(0)}-\sqrt{\tilde D^\chi}$, 
where $l^\chi_0$ is normalized to be $l^{\chi=0}_0=(1, 1, 1)$. 
By substitution Eqs.\ (\ref{rho33}) and (\ref{lprime33}) into Eq.\ (\ref{excessNcurvature}), 
we obtain Eq.\ (\ref{Uinf}).

\end{document}